\documentclass{article}
\usepackage[preprint]{spconf}
\usepackage{amsmath,graphicx}
\usepackage{array}
\usepackage{comment}
\usepackage{multirow}
\usepackage{hyperref}
\usepackage{booktabs}
\usepackage[table,xcdraw]{xcolor}
\renewcommand{\paragraph}[1]{\noindent\textbf{#1}\quad} 

\title{Are E2E ASR models ready for an industrial usage?}
\name{Valentin Vielzeuf, Grigory Antipov}
\address{Orange, 4 rue du Clos Courtel, Cesson-S\'evign\'e, France\\
        \{valentin.vielzeuf,grigory.antipov\}@orange.com}

\begin{document}
\ninept
\maketitle
\begin{abstract}
The Automated Speech Recognition (ASR) community experiences a major turning point with the rise of the fully-neural (End-to-End, E2E) approaches.
At the same time, the conventional hybrid model remains the standard choice for the practical usage of ASR.
According to previous studies, the adoption of E2E ASR in real-world applications was hindered by two main limitations: their ability to generalize on unseen domains and their high operational cost.
In this paper, we investigate both above-mentioned drawbacks by performing a comprehensive multi-domain benchmark of several contemporary E2E models and a hybrid baseline.
Our experiments demonstrate that E2E models are viable alternatives for the hybrid approach, and even outperform the baseline both in accuracy and in operational efficiency.
As a result, our study shows that the generalization and complexity issues are no longer the major obstacle for industrial integration, and draws the community's attention to other potential limitations of the E2E approaches in some specific use-cases.
\end{abstract}
\begin{keywords}
Benchmark, Industry, ASR, E2E
\end{keywords}
\section{Introduction}
\label{sec:intro}
In recent years, Automatic Speech Recognition (ASR) has experienced an impressive breakthrough of performances measured as Word Error Rate (WER), especially on LibriSpeech, the popular academic benchmark of English read speech~\cite{panayotov2015librispeech}.
As testified by the results aggregated by the website PapersWithCode\footnote{\href{https://paperswithcode.com/sota/speech-recognition-on-librispeech-test-clean}{Link to PapersWithCode.}}, the End-To-End (E2E) fully-neural models have significantly outperformed conventional hybrid approaches~\cite{povey2016purely} (the neural part of which is limited to an acoustic model).
The dazzling progress of the E2E models has been mainly due to the proposal of new neural architectures (such as ContextNet~\cite{han2020contextnet} or Conformer~\cite{gulati2020conformer}), to exploitation of large amounts of non-annotated speech data via semi- or self-supervised learning~\cite{xu2020iterative,baevski2020wav2vec}, and to the new data augmentation techniques~\cite{park2019specaugment}.

However, despite the fact that the progress of the E2E models is undeniable, hybrid models still remain a default option when building ASR systems for practical usage. Indeed, recent work highlights at least two major concerns hindering the adoption of such models in an industrial context, namely: (a) their generalization ability, and (b) their computational complexity (and therefore operational costs).

More precisely, several studies~\cite{szymanski2020we, likhomanenko2020rethinking} demonstrate that the scores on academic datasets such as LibriSpeech can be deceptive and poorly generalize on other speech domains.
In particular, Szymanski et al.~\cite{szymanski2020we} urge the community to create new benchmarks, and illustrate that there is a huge gap between the WERs measured on popular academic datasets, and the WERs measured on private ones for various real-life use-cases.
In the same spirit, Likhomanenko et al.~\cite{likhomanenko2020rethinking} show that there is little generalization between performances of the contemporary E2E ASR models across public benchmark datasets, and that the models trained on LibriSpeech particularly struggle to transfer to other domains.
Thus, a major milestone to better quantify this lack of generalization consists in building comprehensive evaluation datasets composed of speech of various nature (\textit{i.e.} multi-domain evaluation). This is in line with very recent work proposing to aggregate different existing datasets~\cite{evain2021lebenchmark}.
The mentioned studies clearly identify the generalization problem and Aksenova et al.~\cite{aksenova2021might} following some earlier work~\cite{amodei2016deep,narayanan2018toward,kanda2021large} propose to address it by diversifying the training datasets (\textit{i.e.} multi-domain ASR training).

On the other hand, E2E models are often associated with a larger computational burden. For instance, recent models~\cite{baevski2020wav2vec} reach 300M parameters which represents around 30 times as much as the size of the acoustic part of the traditional hybrid models often used in industry~\cite{povey2016purely}. This strongly motivates the community to focus on the reduction of the computational cost of the E2E ASR approaches.
For example, there is a strong interest in methods allowing online E2E ASR decoding without latency~\cite{yu2020dual,yu2021fastemit}.
Another research direction targets lighter architectures which may help to reduce both training and inference time in several use-cases. Indeed, optimized convolutional models have been proposed using the depthwise convolution and the simple CTC loss~\cite{pratap2019wav2letter++,majumdar2021citrinet}.
Transformer-based models have also been studied in depth. For instance, several ``efficient'' Conformers~\cite{wang2021efficient,li2021efficient,burchi2021efficient} have been proposed recently.

Summarizing, the above-mentioned industrial constraints face the community with a trade-off between a high and reliable ASR accuracy and a low resource consumption (in the spirit of the Occam's Razor principle).
In this paper, we demonstrate that there are contemporary E2E models which perfectly match the presented compromise and, therefore, show that the generalization and efficiency are no longer the major barrier to the industrial adoption of the E2E models.
To this end, we benchmark promising E2E architectures comparing with a standard hybrid ASR model by
(a) performing both training and evaluation in a multi-domain context; and 
(b) measuring both the accuracy and the efficiency in a real-world aware manner.


\section{Multi-domain E2E vs. hybrid benchmark}
\label{sec:methodo}

\begin{table*}[ht]
\centering
\begin{tabular}{cc|cccccc|c|c|cc|c} 
\toprule
\multicolumn{2}{c|}{\textbf{\textit{Models}}}                                                                     & \multicolumn{7}{c|}{\multirow{2}{*}{\begin{tabular}[c]{@{}c@{}}\textbf{\textit{Mutli-domain Accuracy }}\\\textbf{\textit{(WER in \%)}}\end{tabular}}} & \multicolumn{4}{c}{\textit{\textbf{Computational Cost}}}                                                                                                                                                                                           \\ 
\cline{10-13}
\multicolumn{2}{l|}{}                                                                                             & \multicolumn{7}{c|}{}                                                                                                                                 & \multirow{2}{*}{\begin{tabular}[c]{@{}c@{}}\textbf{Training }\\\textbf{time (days)}\end{tabular}} & \multicolumn{2}{c|}{\textbf{iRTF}}                 & \multirow{2}{*}{\begin{tabular}[c]{@{}c@{}}\textbf{\#}\\\textbf{params}\end{tabular}}  \\ 
\cline{3-9}\cline{11-12}
\multicolumn{2}{l|}{}                                                                                             & \textbf{FR}   & \textbf{LS\_c} & \textbf{LS\_o} & \textbf{SB}   & \textbf{TED} & \textbf{WSJ} & \textbf{Overall}                                      &                                                                                                   & \textbf{CPU}                 & \textbf{GPU}        &                                                                                        \\ 
\hline
\multicolumn{2}{c|}{\textbf{Hybrid}}                                                                              & 37.2          & 11.0           & 25.4           & 25.3          & 12.1         & 9.3          & 20.0                                                  & 7                                                                                                 & 2                            & N/A                 & 8M                                                                                   \\ 
\hline
\multirow{2}{*}{\begin{tabular}[c]{@{}c@{}}\textbf{Conformer}\\\textbf{ (small)}\end{tabular}}  & \textbf{Greedy} & 30.1          & 5.8            & 13.7           & 20.1          & 9.4          & 6.7          & 14.3                                                  & \multirow{2}{*}{7}                                                                                & \multirow{2}{*}{\textbf{33}} & \multirow{2}{*}{50} & \multirow{2}{*}{13M}                                                                   \\ 
\cline{2-9}
                                                                                                & \textbf{+ LM}   & 27.0          & 4.9            & 11.8           & 18.7          & 8.1          & 5.5          & 12.6                                                  &                                                                                                   &                              &                     &                                                                                        \\ 
\hline
\multirow{2}{*}{\begin{tabular}[c]{@{}c@{}}\textbf{Conformer}\\\textbf{ (medium)}\end{tabular}} & \textbf{Greedy} & 32.3          & 6.2            & 13.9           & 21.2          & 9.9          & 6.8          & 15.1                                                  & \multirow{2}{*}{7}                                                                                & \multirow{2}{*}{17}          & \multirow{2}{*}{50} & \multirow{2}{*}{30M}                                                                   \\ 
\cline{2-9}
                                                                                                & \textbf{+ LM}   & 28.0          & 4.9            & 11.7           & 19.1          & 8.1          & 5.5          & 12.9                                                  &                                                                                                   &                              &                     &                                                                                        \\ 
\hline
\multirow{2}{*}{\begin{tabular}[c]{@{}c@{}}\textbf{Citrinet}\\\textbf{ (small)}\end{tabular}}   & \textbf{Greedy} & 33.8          & 5.0            & 12.9           & 22.1          & 9.2          & 6.5          & 14.9                                                  & \multirow{2}{*}{7}                                                                                & \multirow{2}{*}{25}          & \multirow{2}{*}{50} & \multirow{2}{*}{10M}                                                                  \\ 
\cline{2-9}
                                                                                                & \textbf{+ LM}   & 32.0          & 4.4            & 11.4           & 21.2          & 8.1          & 5.1          & 13.7                                                  &                                                                                                   &                              &                     &                                                                                        \\ 
\hline
\multirow{2}{*}{\begin{tabular}[c]{@{}c@{}}\textbf{Citrinet}\\\textbf{ (medium)}\end{tabular}}  & \textbf{Greedy} & 28.6          & \textbf{4.0}   & \textbf{10.2}  & 19.3          & \textbf{7.6} & \textbf{5.0} & \textbf{12.5}                                         & \multirow{2}{*}{7}                                                                                & \multirow{2}{*}{10}          & \multirow{2}{*}{50} & \multirow{2}{*}{21M}                                                                   \\ 
\cline{2-9}
                                                                                                & \textbf{+ LM}   & 26.3          & 6.5            & 10.6           & 19.7          & 7.7          & 8.6          & 13.2                                                  &                                                                                                   &                              &                     &                                                                                        \\ 
\hline
\multirow{2}{*}{\textbf{CRDNN}}                                                                 & \textbf{Greedy} & \textbf{25.5} & 5.5            & 15.1           & \textbf{18.1} & 8.3          & \textbf{5.1} & 12.9                                                  & \multirow{2}{*}{14}                                                                               & \multirow{2}{*}{2.5}         & \multirow{2}{*}{50} & \multirow{2}{*}{120M}                                                                  \\ 
\cline{2-9}
                                                                                                & \textbf{+ LM}   & 27.2          & 7.0            & 17.2           & 21.8          & 11.0         & 6.3          & 15.1                                                  &                                                                                                   &                              &                     &                                                                                        \\
\bottomrule
\end{tabular}
\caption{Summary of the principal benchmark results: Multi-domain Accuracy and Computational Cost. ASR WERs for all compared models are provided on the datasets described in Subsection~\ref{subsec:datasets_definition}, namely: Franglish (FR), LibriSpeech-clean (LS\_c), LibriSpeech-other (LS\_o), SwitchBoard (SB), TED (TED-LIUM), and WSJ. The column ``Overall'' is the average of the 6 scores. For each compared model, the WER scores are provided for the greedy and language model (LM)-based decoding. The latter is performed with a $3$-gram LM and a beam size of $4$. The training times are measured on contemporary work stations equipped with $4$ Nvidia 2080 Ti GPUs. The reported inverted RTFs (iRTF) correspond to the average iRTFs over the test dataset which are measured in virtual machines where the compared models are run either on a single CPU or GPU (Nvidia 2080 Ti). The sizes of the models are reported in the last column.}
\label{tab:main_results_table}
\end{table*}

\subsection{Benchmark dataset}
\label{subsec:datasets_definition}

\paragraph{Multi-domain training \& evaluation} 
We follow the recommendations of the recent studies mentioned in Section~\ref{sec:intro} by designing a multi-domain dataset for our benchmark.
We choose to work in English because it is the language with the largest choice of public ASR datasets. 
More precisely, we construct a collection of datasets issued from various application domains, namely: read speech (LibriSpeech~\cite{panayotov2015librispeech}, 870h of speech, 85K of unique subwords), phone conversations (SwitchBoard~\cite{SwitchBoard1993}, 250h of speech, 26K of unique subwords), dictation (WSJ~\cite{WSJ1993}, 65h of speech, 14K of unique subwords), prepared talks (TED-LIUM~\cite{hernandez2018specom}, 250h of speech, 44K of unique subwords) and spontaneous non-native speech (we use a private dataset called Franglish, composed of the meeting recordings of French natives speaking English in a spontaneous manner, 16h of speech, 6K of unique subwords).
Thus, our collection is composed of $4$ public well-studied datasets of various nature and of a private dataset.
The latter (Franglish) is added to the benchmark as, to the best of our knowledge, there are no public datasets with non-native spontaneous English speech.
At the same time, such speech includes grammatical imperfections as well as hesitations, repeated words and interrupted phrases, augments the acoustic variability of the data, and therefore, represents a real challenge for ASR systems (which perfectly fits our objective of evaluating the generalization of the compared models).

\paragraph{Data preparation}
The multi-domain datasets are
normalized with the help of the Nemo Text Normalization tool~\cite{NeMoTextNorm}.
For simplicity and following previous studies, we leave Voice Activity Detection (VAD) out of the main scope of the present benchmark by employing the ground-truth (oracle) speech segments both for training and evaluation.
We motivate and confirm this choice by a dedicated ablation study which is presented in Subsection~\ref{subsec:VAD_impact}.
Finally, the training / valid / test splits of the public datasets of our collection follow the respective protocols established in the community.
Franglish dataset is randomly split in proportion $80$\% / $10$\% / $10$\%.

\subsection{Compared ASR models}
\label{subsec:compared_ASR_models}

\paragraph{Hybrid system}
In this benchmark, we use a standard hybrid approach from~\cite{povey2016purely} as a baseline for the evaluated ASR models.
Roughly speaking, it consists of $2$ parts: an Acoustic Model (AM) and a Language Model (LM).
AM is a TDNN~\cite{peddinti2015time} which is used to predict a posterior distribution over the tied Hidden Markov Model (HMM) states corresponding to context-dependent phonemes (biphones).
These posterior distributions are then combined with a lexicon and a n-gram LM in order to construct a search graph in a form of WFST~\cite{mohri2009weighted}.
During the inference, the decoding is done via the beam search which looks for the best paths in the constructed graph.

\paragraph{E2E models}
Obviously, it is infeasible to evaluate all SOTA ASR models in the frame of one benchmark.
Therefore, we select several SOTA-level representatives of the $3$ large families of E2E models, namely: the recurrent ones, the fully-convolutional ones and the Transformer-based ones.
More precisely, hereafter, we present the selected \textit{encoder} architectures, while the same CTC decoder~\cite{graves2006connectionist} is used for all E2E ASR models in the present benchmark.

\textit{Recurrent ASR encoder}.
Being natural candidates for modelling sequential data (such as speech recordings), RNN-based models are notorious for their computational complexity.
For this benchmark, we choose a popular \textbf{CRDNN} architecture, which is a combination of a CNN, a RNN and a MLP composed of $120$M trainable weights.
In particular, we use the SpeechBrain's~\cite{speechbrain} implementation of this architecture~\footnote{\href{https://speechbrain.readthedocs.io/en/latest/API/speechbrain.lobes.models.CRDNN.html}{Link to the CRDNN model description.}}.

\textit{Fully-convolutional ASR encoder} is another promising family of models composed of convolutional blocks, which allow faster training and inference comparing to recurrent models.
Citrinet~\cite{majumdar2021citrinet} is SOTA-level example of the family.
We evaluate $2$ versions of Citrinet: \textbf{Citrinet-small} and \textbf{Citrinet-medium} ($10$M and $30$M parameters, respectively).
 
\textit{Transformer-based ASR encoder}.
Transformer-based models are the SOTA today.
But their computationally cost is high due to the quadratic complexity of the self-attention mechanism w.r.t. the input size.
In this benchmark, we employ the Conformer~\cite{gulati2020conformer} 
encoder combining self-attention and convolutional layers.
In particular, we evaluate two versions of this architecture of varying complexity, namely: \textbf{Conformer-small} and \textbf{Conformer-medium} ($13$M and $30$M parameters, respectively).


\subsection{Evaluated metrics}

\paragraph{Accuracy}
WER (defined as the ratio between the sum of the substitution $S$, deletion $D$ and insertion $I$ errors and the total number of words $N$ in the ground-truth transcription: $WER=\frac{S+D+I}{N}$) is by far the most widely adopted metric for evaluation of the ASR systems.
Therefore, we use it in our benchmark for evaluation of the multi-domain accuracy.

\paragraph{Computational cost}
In addition to the number of parameters, the compared ASR models are evaluated according to $3$ criteria which are particularly important for the integration of ASR systems, namely: the training time, the inference time and the required RAM.
We allocate the same training time budget of $7$ days of calculation on a modern workstation equipped with $4$ Nvidia 2080 Ti GPUs for all E2E models.
This value corresponds to the time required by the baseline hybrid model for convergence on the selected collection of training datasets.
We make a single exception to this training budget for the CRDNN model which (due to its computational complexity) requires at least twice as much time to converge to competitive ASR performances.
For the inference time, we employ the popular inverted Real Time Factor (iRTF) metric measuring the ratio between the real time of the input recording and the time spent by an ASR system for its transcription.
Moreover, the inference time and the memory requirements obviously depend on the duration of the input audio recordings.
Therefore, in our benchmark, we also evaluate both criteria by varying the size of inputs in order to evaluate the scalability of the compared ASR systems.

\section{Benchmark results}
\label{sec:benchmarking}
\begin{figure*}[ht]
\centering
\begin{minipage}{0.47\linewidth}
\centering
  \includegraphics[width=0.95\linewidth]{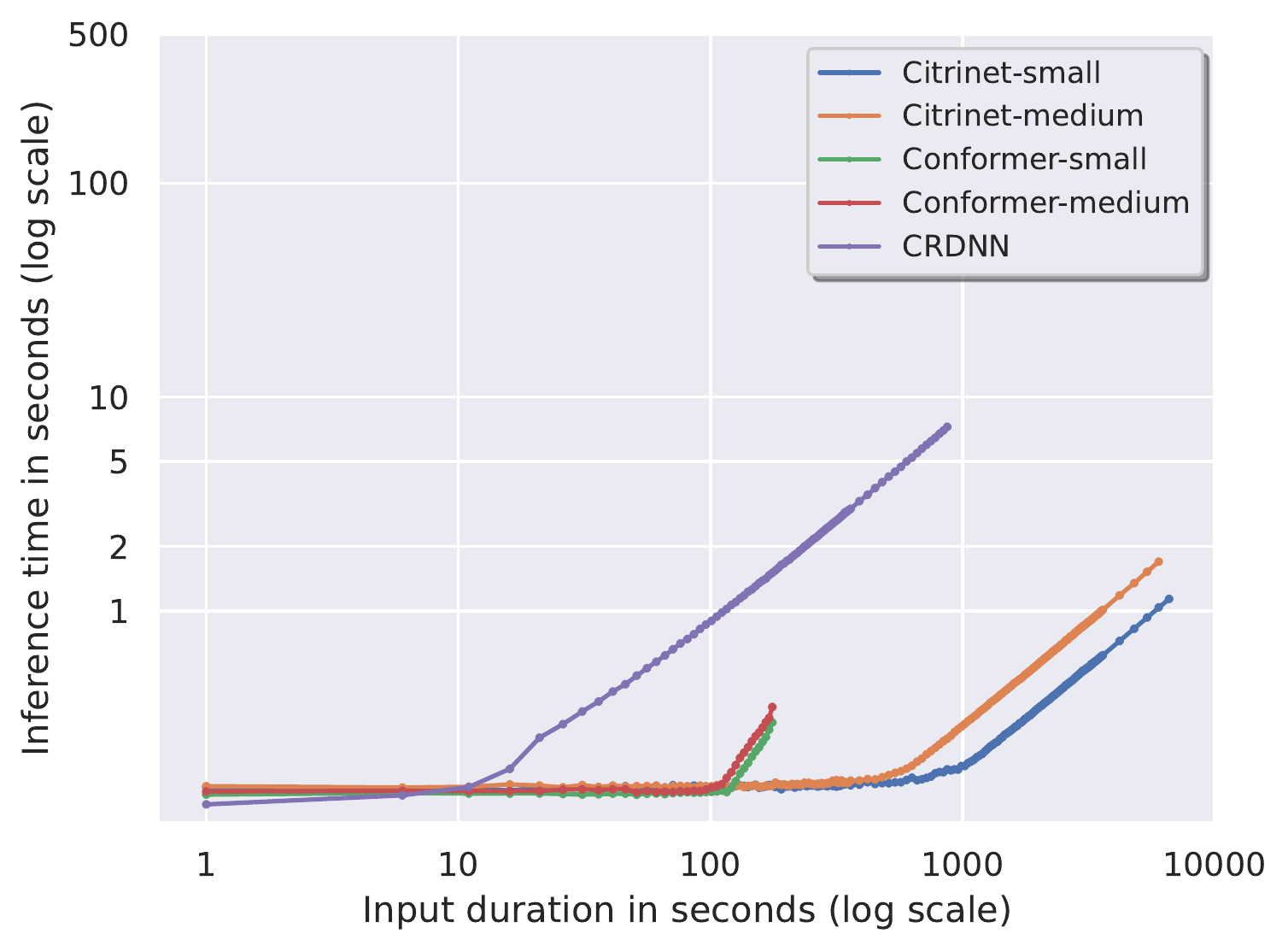}
\textbf{(a)}
\end{minipage}
\begin{minipage}{0.47\linewidth}
\centering
  \includegraphics[width=\linewidth]{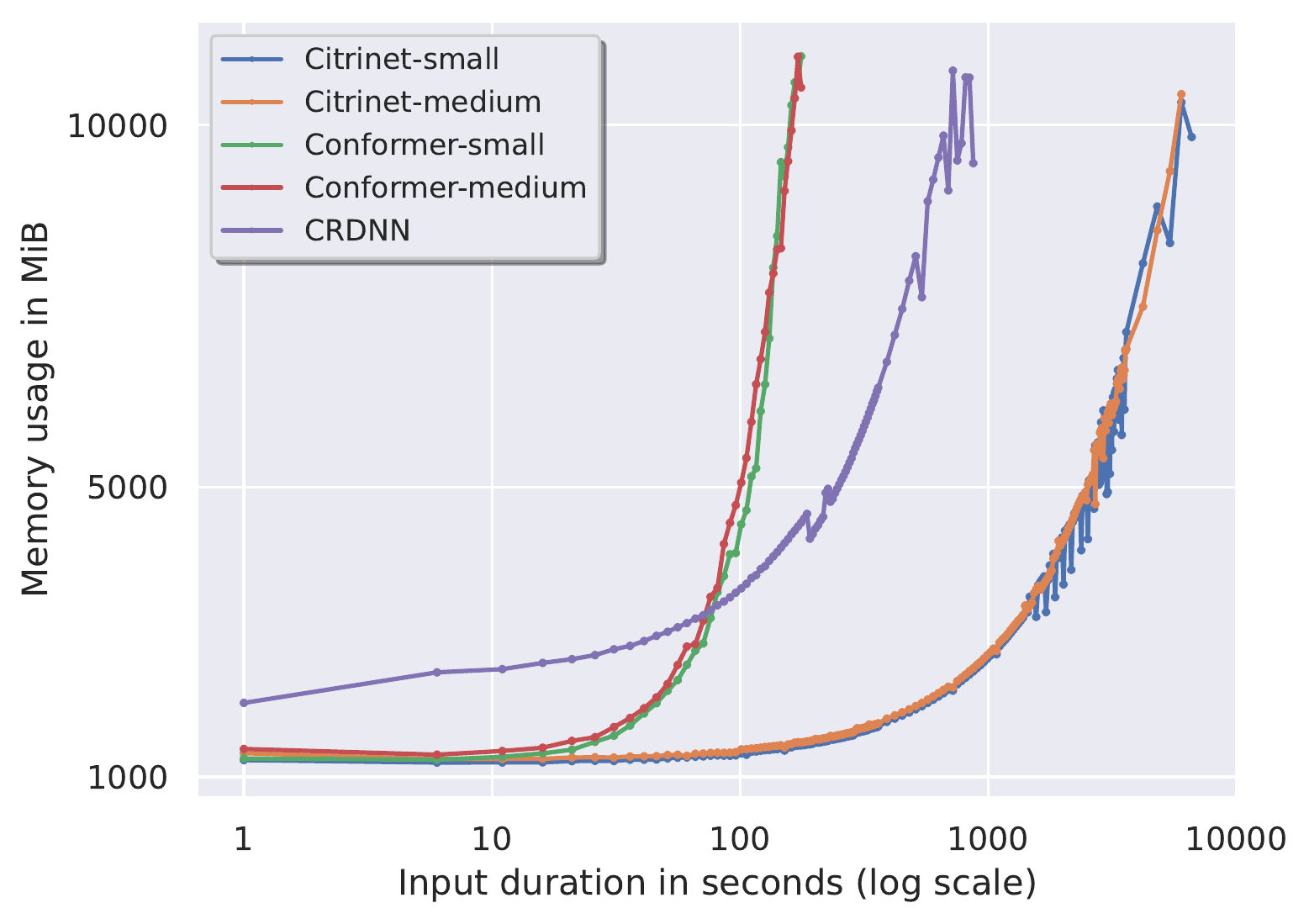}
\textbf{(b)}
\end{minipage}
\caption{Greedy inference time \textbf{(a)} and memory usage \textbf{(b)} vs. the input recording's length measured for the compared ASR E2E models on a contemporary workstation equipped with a single Nvidia 2080 GPU Ti.}
\label{fig:time}
\end{figure*}
\subsection{Multi-Domain accuracy}
The principal results of our benchmark are summarized in Table~\ref{tab:main_results_table}.
The WER scores significantly vary depending on the evaluation dataset (and hence, on the target domain) which corroborates with the previous studies discussed in Section~\ref{sec:intro}.
Expectably, the best transcription results are witnessed on the read speech (from $4$\% to $11$\% of WER on LibriSpeech\_clean) which is widely recognized as the easiest ASR use-case.
The results on the 16kHz-sampled prepared speech are close to those of the read speech (from $5$\% to $12$\% on TED-LIUM and WSJ).
On the contrary, the accuracy drastically drops on the 8kHz-sampled phone speech (from $18$\% to $25$\% on SwitchBoard) and, above all, on the spontaneous accented speech (from $25$\% to $37$\% on Franglish) which clearly represents the biggest challenge among the domains included in the benchmark.

The results in Table~\ref{tab:main_results_table} are unequivocal in terms of the ASR accuracy comparison between the E2E and hybrid models.
Indeed, as one may observe, \textit{all} compared E2E ASR models outperform the hybrid one by a large margin on \textit{all} evaluation datasets.
The relative WER improvements brought by the E2E models w.r.t. the hybrid one vary from about $30$\% to $65$\% depending on the dataset.
In other words, the superiority of the E2E models does not limit to LibriSpeech, but is a rather general tendency on all ASR use-cases.

When comparing the E2E models between each other, the ``Overall'' column of Table~\ref{tab:main_results_table} demonstrates that the Citrinets slightly outperform the Conformers of the similar sizes.
However, it should be noted that (as explained in Subsection~\ref{subsec:production_cost}) we set the same training time budget of $7$ days for all E2E models except for CRDNN.
And the Conformers appear to converge a little slower than the Citrinets.
Therefore, we suspect that the Conformers might be slightly underfit, which partially explains the worse accuracies than those obtained by the Citrinets.
This is particularly true for the Conformer-medium which is outperformed by its simpler (but better converged) Conformer-small counterpart.
At last, CRDNN obtains excellent WERs which are on-par with the ones of Citrinet-medium, but, as discussed in Subsection~\ref{subsec:production_cost}, at much greater cost than the latter.

Finally, it is worth noting that all compared E2E models perform reasonably well even with the simplest greedy decoding.
This confirms that E2E models do not exclusively focus on the acoustics of the speech (as it is the case for the acoustic neural network of the hybrid model), but rather jointly learn acoustic and linguistics aspects of the language.
Moreover, the WER scores of the most accurate E2E models, namely Citrinet-medium and CRDNN, are even deteriorated by the added LM.
This can be explained by the fact that a simple $3$-gram LM is used in our experiments, and probably Citrinet-medium and CRDNN implicitly learn a better representation of the language than the one provided by such a trivial LM.

\begin{table}[t]
\centering
\begin{tabular}{c|c|c|c} 
\hline
\multirow{2}{*}{\begin{tabular}[c]{@{}c@{}}\textit{\textbf{VAD}}\\\end{tabular}} & \multirow{2}{*}{\textit{\textbf{WER (in \%)}}} & \multicolumn{2}{c}{\textit{\textbf{iRTF}}}     \\ 
\cline{3-4}
                                                                                 &                                                & \textit{\textbf{CPU}} & \textit{\textbf{GPU}}  \\ 
\hline
\textit{\textbf{None}}                                                           & 14.5                                           & N/A                   & N/A                    \\ 
\hline
\textit{\textbf{Oracle}}                                                         & 14.3                                           & N/A                   & N/A                    \\ 
\hline
\textit{\textbf{Energy-based}}                                                   & 14.3                                           & 68                   & N/A                    \\ 
\hline
\textit{\textbf{SpeechBrain}}                                                    & 14.7                                           & 104                   & 99                     \\ 
\hline
\textit{\textbf{Nemo}}                                                           & 14.7                                           & 19                    & 222                    \\ 
\hline
\textit{\textbf{Silero}}                                                         & 17.0                                           & 56                    & 55                     \\
\hline
\end{tabular}
\caption{Ablation study of VAD for E2E ASR. Various VAD approaches are compared w.r.t. their impact on the resulting transcription accuracies (WER) and the respective computational costs (iRTF).
The WER scores are averaged over all evaluation datasets. All results are reported for Conformer (small).}
\label{tab:VAD_ablation}
\end{table}

\subsection{Computational Cost}
\label{subsec:production_cost}
Today, the average cost of an hour rental of a workstaion with $4$ contemporary GPUs is around \$ $2$.
Thus, an average training run costs \$ $336$ for the Conformers, Citrinets, and for the hybrid system and twice as much (\textit{i.e.} \$ $672$) for the CRDNN.
The WER improvements brought by CRDNN which are reported in Table~\ref{tab:main_results_table} seem marginal comparing to its training cost.
Also, one should keep in mind that several training runs are often needed to maintain the model with the most recent data or to adapt it to specific use-cases, multiplying the original cost.

When the model is finally trained and delivered for industrial usage, the main concern is its inference cost. In other words, how long does it take to process one standard input with a given hardware?
From Table~\ref{tab:main_results_table}, we can see that all chosen models are faster than real-time, even using only one modern CPU. Yet, CRDNN and the hybrid system are only $2$ times faster than real-time, while Conformer-small is $33$ times faster, meaning that Conformer-small would need about $16$ times less resources to guarantee the same rapidity of the transcription as the one provided by the hybrid model.
Given an everyday intensive usage, such enormous difference may represent a very large cost gain arguing in favor of the deployment of light E2E models such as Conformer-small or Citrinet-small.

One may notice that according to the results in Table~\ref{tab:main_results_table}, GPUs do not seem to accelerate the inference of the E2E ASR models.
However, this is only due to our evaluation protocol, which processes input sequences one by one and not in batches.
In order to quantify the potential benefits brought by the GPU usage at inference, batches or longer sequences should be fed to the model.
Hence, in Figure~\ref{fig:time}-(a) we extend the experiment to longer sequences.
One may observe that for very short sequences (less than $10$ seconds), it's difficult to compare the E2E models, as the GPU is used sub-optimally.
For longer sequences, the difference between CRDNN and the other E2E models becomes obvious, the former being much slower (though keeping a very decent iRTF).
The longest sequences perfectly illustrate that the Conformers are slower than the Citrinets, which is due to the squared complexity of the self-attention w.r.t. the input duration.
At last, one may observe an outstanding result of Citrinet-small which manages to process an hour of speech in less than a second. 

The number of parameters is another important concern, as it is directly connected to the minimal required disk storage.
For instance CRDNN is around $10$ times larger than the smallest E2E models and therefore, does not imply the same industrial constraints.
Concerning the specific case of the hybrid model, the reported parameters only include the acoustic neural network part and therefore, do not represent the real storage requirements.
Moreover, for the E2E models, Figure~\ref{fig:time}-(b) shows that the problem is not only about the number of parameters, but also about the dependency between the memory usage and the length of the processed inputs.
Indeed, the Conformers introduce an enormous memory burden when the sequences are too long, quickly saturating the GPU storage, while even the largest convolutional or recurrent models like the CRDNN are able to process such sequences while fitting on a modern GPU.

\subsection{Impact of VAD}
\label{subsec:VAD_impact}

As mentioned in Subsection~\ref{subsec:datasets_definition}, all results discussed above and summarized in Table~\ref{tab:main_results_table} have been obtained with a ground-truth speech segmentation leaving VAD out of the scope of the main study.
Below, we motivate this choice by comparing alternative VAD approaches with the Conformer (small) E2E architecture.
More precisely, Table~\ref{tab:VAD_ablation} reports the results of an ablation study where Conformer (small) transcribes the speech recordings from our benchmark evaluation datasets segmented according to one of the following VAD methods: ``None'' (no segmentation at all); ``Oracle'' (ground-truth segmentation); ``Energy-based'' (segmentation is based on the amplitudes of the input waves); and $3$ open-source off-the-shelf VAD solutions, namely: SpeechBrain~\cite{speechbrain}, Nemo~\cite{kuchaiev2019nemo} and Silero~\cite{Silero_VAD}.
As one can witness from the WER scores in Table~\ref{tab:VAD_ablation}, the impact of VAD on the accuracy of the speech recognition is extremely limited.
Indeed, the transcriptions generated from the ground-truth segmentations are barely ($0.2$ WER points) better than the ones generated from the raw, non-segmented speech recordings.
In average, the absence of segmentation works even better than using the off-the-shelf VAD approaches (especially, the Silero's one).
All in all, the performed ablation perfectly illustrates that during the E2E training, ASR models implicitly learn to distinguish the speech from the noise.
This positively distinguishes E2E models from the hybrid ones as the latter are notorious for being dependent on the quality of the speech segmentation.

\section{Discussion and Conclusion}
\label{sec:conclusion}
In this work, we have proposed a multi-domain training and evaluation benchmark and studied 2 reported practical limitations of E2E ASR models: their generalization ability and computational cost.
The evaluated E2E models have consistently outperformed the strong hybrid baseline system in terms of the multi-domain WER.
The estimated computational costs also testify in favor of the E2E models. 
Indeed, all evaluated E2E models (except for CRDNN) significantly reduce both training time and inference costs w.r.t. the hybrid approach. 
We have also shown that E2E models scale well w.r.t. the input recordings duration (processing up to one hour at once for the Citrinets).
As a result, our experiments demonstrate that generalization and efficiency can no longer be considered as central issues preventing the industrial usage of E2E ASR, which allows us to positively answer the question put in the paper's title. 

As a side result, the benchmark has pointed the Citrinets as a better trade-off (than the Conformers and CRDNN) between the resulting ASR accuracy and the training / inference complexity, and that a LM-free greedy decoding is sufficient to obtain decent performances on all tested use-cases.
Another interesting side observation of our study is that E2E ASR models are not really sensitive to the quality of the input speech segmentation which is an important advantage w.r.t. the hybrid models.

Finally, it is important to notice that all E2E experiments have been done with a trivial 3-gram LM.
Thus, the study on the impact of complex LM integration on the E2E model's accuracy and efficiency constitutes an important direction for future work.
Another promising path of research would consist in further extending the evaluation protocol in order to assess the models' adaptability.
Indeed, the hybrid ASR systems are known to be easily adaptable to a new lexicon, but can we say likewise regarding the E2E ones?

\vfill
\pagebreak
\bibliographystyle{IEEEbib}
\bibliography{strings,refs}

\begin{thebibliography}{10}

\bibitem{panayotov2015librispeech}
Vassil Panayotov, Guoguo Chen, Daniel Povey, et~al.,
\newblock ``Librispeech: an asr corpus based on public domain audio books,''
\newblock in {\em ICASSP}, 2015.

\bibitem{povey2016purely}
Daniel Povey, Vijayaditya Peddinti, Daniel Galvez, et~al.,
\newblock ``Purely sequence-trained neural networks for asr based on
  lattice-free mmi.,''
\newblock in {\em Interspeech}, 2016.

\bibitem{han2020contextnet}
Wei Han, Zhengdong Zhang, Yu~Zhang, et~al.,
\newblock ``Contextnet: Improving convolutional neural networks for automatic
  speech recognition with global context,''
\newblock in {\em Interspeech}, 2020.

\bibitem{gulati2020conformer}
Anmol Gulati, James Qin, Chung-Cheng Chiu, et~al.,
\newblock ``Conformer: Convolution-augmented transformer for speech
  recognition,''
\newblock in {\em Interspeech}, 2020.

\bibitem{xu2020iterative}
Qiantong Xu, Tatiana Likhomanenko, Jacob Kahn, et~al.,
\newblock ``Iterative pseudo-labeling for speech recognition,''
\newblock in {\em Interspeech}, 2020.

\bibitem{baevski2020wav2vec}
Alexei Baevski, Henry Zhou, Abdelrahman Mohamed, et~al.,
\newblock ``wav2vec 2.0: A framework for self-supervised learning of speech
  representations,''
\newblock in {\em NeurIPS}, 2020.

\bibitem{park2019specaugment}
Daniel~S Park, William Chan, Yu~Zhang, et~al.,
\newblock ``Specaugment: A simple data augmentation method for automatic speech
  recognition,''
\newblock in {\em Interspeech}, 2019.

\bibitem{szymanski2020we}
Piotr Szyma{\'n}ski, Piotr {\.Z}elasko, Mikolaj Morzy, et~al.,
\newblock ``Wer we are and wer we think we are,''
\newblock in {\em EMNLP}, 2020.

\bibitem{likhomanenko2020rethinking}
Tatiana Likhomanenko, Qiantong Xu, Vineel Pratap, et~al.,
\newblock ``Rethinking evaluation in asr: Are our models robust enough?,''
\newblock {\em arXiv preprint arXiv:2010.11745}, 2020.

\bibitem{evain2021lebenchmark}
Solene Evain, Ha~Nguyen, Hang Le, et~al.,
\newblock ``Lebenchmark: A reproducible framework for assessing self-supervised
  representation learning from speech,''
\newblock {\em arXiv preprint arXiv:2104.11462}, 2021.

\bibitem{aksenova2021might}
Al{\"e}na Aks{\"e}nova, Daan van Esch, James Flynn, et~al.,
\newblock ``How might we create better benchmarks for speech recognition?,''
\newblock in {\em Workshop on Benchmarking}, 2021.

\bibitem{amodei2016deep}
Dario Amodei, Sundaram Ananthanarayanan, Rishita Anubhai, et~al.,
\newblock ``Deep speech 2: End-to-end speech recognition in english and
  mandarin,''
\newblock in {\em ICML}, 2016.

\bibitem{narayanan2018toward}
Arun Narayanan, Ananya Misra, Khe~Chai Sim, et~al.,
\newblock ``Toward domain-invariant speech recognition via large scale
  training,''
\newblock in {\em SLT}, 2018.

\bibitem{kanda2021large}
Naoyuki Kanda, Guoli Ye, Yu~Wu, et~al.,
\newblock ``Large-scale pre-training of end-to-end multi-talker asr for meeting
  transcription with single distant microphone,''
\newblock {\em arXiv preprint arXiv:2103.16776}, 2021.

\bibitem{yu2020dual}
Jiahui Yu, Wei Han, Anmol Gulati, et~al.,
\newblock ``Dual-mode asr: Unify and improve streaming asr with full-context
  modeling,''
\newblock in {\em ICLR}, 2020.

\bibitem{yu2021fastemit}
Jiahui Yu, Chung-Cheng Chiu, Bo~Li, et~al.,
\newblock ``Fastemit: Low-latency streaming asr with sequence-level emission
  regularization,''
\newblock in {\em ICASSP}, 2021.

\bibitem{pratap2019wav2letter++}
Vineel Pratap, Awni Hannun, Qiantong Xu, et~al.,
\newblock ``Wav2letter++: A fast open-source speech recognition system,''
\newblock in {\em ICASSP}, 2019.

\bibitem{majumdar2021citrinet}
Somshubra Majumdar, Jagadeesh Balam, Oleksii Hrinchuk, et~al.,
\newblock ``Citrinet: Closing the gap between non-autoregressive and
  autoregressive end-to-end models for automatic speech recognition,''
\newblock {\em arXiv preprint arXiv:2104.01721}, 2021.

\bibitem{wang2021efficient}
Xiong Wang, Sining Sun, Lei Xie, et~al.,
\newblock ``Efficient conformer with prob-sparse attention mechanism for
  end-to-endspeech recognition,''
\newblock {\em arXiv preprint arXiv:2106.09236}, 2021.

\bibitem{li2021efficient}
Shengqiang Li, Menglong Xu, and Xiao-Lei Zhang,
\newblock ``Efficient conformer-based speech recognition with linear
  attention,''
\newblock {\em arXiv preprint arXiv:2104.06865}, 2021.

\bibitem{burchi2021efficient}
Maxime Burchi and Valentin Vielzeuf,
\newblock ``Efficient conformer: Progressive downsampling and grouped attention
  for automatic speech recognition,''
\newblock in {\em ASRU}, 2021.

\bibitem{SwitchBoard1993}
John~J. Godfrey and Edward Holliman,
\newblock ``Switchboard-1 release 2,'' Linguistic Data Consortium, 1993.

\bibitem{WSJ1993}
John~S. Garofolo, David Graff, Doug Paul, et~al.,
\newblock ``Csr-i (wsj0) complete,'' Linguistic Data Consortium, 1993.

\bibitem{hernandez2018specom}
François Hernandez, Vincent Nguyen, Sahar Ghannay, et~al.,
\newblock ``Ted-lium 3: twice as much data and corpus repartition for
  experiments on speaker adaptation,''
\newblock in {\em SPECOM}, 2018.

\bibitem{NeMoTextNorm}
Yang Zhang, Evelina Bakhturina, Kyle Gorman, et~al.,
\newblock ``Nemo inverse text normalization: From development to production,''
\newblock {\em arXiv preprint arXiv:2104.05055}, 2021.

\bibitem{peddinti2015time}
Vijayaditya Peddinti, Daniel Povey, and Sanjeev Khudanpur,
\newblock ``A time delay neural network architecture for efficient modeling of
  long temporal contexts,''
\newblock in {\em Interspeech}, 2015.

\bibitem{mohri2009weighted}
Mehryar Mohri,
\newblock ``Weighted automata algorithms,''
\newblock in {\em Handbook of weighted automata}, 2009.

\bibitem{graves2006connectionist}
Alex Graves, Santiago Fern{\'a}ndez, Faustino Gomez, et~al.,
\newblock ``Connectionist temporal classification: labelling unsegmented
  sequence data with recurrent neural networks,''
\newblock in {\em ICML}, 2006.

\bibitem{speechbrain}
Mirco Ravanelli, Titouan Parcollet, Peter Plantinga, et~al.,
\newblock ``{SpeechBrain}: A general-purpose speech toolkit,'' 2021,
\newblock arXiv:2106.04624.

\bibitem{kuchaiev2019nemo}
Oleksii Kuchaiev, Jason Li, Huyen Nguyen, Oleksii Hrinchuk, Ryan Leary, Boris
  Ginsburg, Samuel Kriman, Stanislav Beliaev, Vitaly Lavrukhin, Jack Cook,
  et~al.,
\newblock ``Nemo: a toolkit for building ai applications using neural
  modules,''
\newblock {\em arXiv preprint arXiv:1909.09577}, 2019.

\bibitem{Silero_VAD}
Silero Team,
\newblock ``Silero vad: pre-trained enterprise-grade voice activity detector
  (vad), number detector and language classifier,''
  \url{https://github.com/snakers4/silero-vad}, 2021.

\end{thebibliography}

\end{document}